\documentclass[conference]{IEEEtran}
\usepackage{cite}
\usepackage[squaren, Gray, cdot]{SIunits}
\usepackage[usenames,dvipsnames,svgnames,table]{xcolor}

\ifCLASSINFOpdf
   \usepackage[pdftex]{graphicx}
\else
\fi
\usepackage{amsmath}

\usepackage{array}

\begin{document}
\title{ Low Power In-Memory Implementation \\  of  Ternary Neural Networks  with \\ Resistive RAM-Based Synapse
\thanks{\textbf{This work was supported by the ERC Grant NANOINFER (715872) and  the ANR grant NEURONIC (ANR-18-CE24-0009).}}
}


\author{\IEEEauthorblockN{
A. Laborieux\IEEEauthorrefmark{1}, 
M. Bocquet\IEEEauthorrefmark{2}, 
T. Hirtzlin\IEEEauthorrefmark{1}, 
J.-O. Klein\IEEEauthorrefmark{1}, L. Herrera Diez\IEEEauthorrefmark{1}, E. Nowak\IEEEauthorrefmark{3}, \\ E. Vianello\IEEEauthorrefmark{3}, J.-M. Portal\IEEEauthorrefmark{2} and D. Querlioz\IEEEauthorrefmark{1}}
\IEEEauthorblockA{\IEEEauthorrefmark{1}Universit\'e Paris-Saclay, CNRS, C2N, 91120 Palaiseau, France.  Email: damien.querlioz@c2n.upsaclay.fr}
\IEEEauthorblockA{\IEEEauthorrefmark{2}
IM2NP,
Univ. Aix-Marseille et Toulon, CNRS, France.
\IEEEauthorrefmark{3}CEA, LETI, Grenoble, France.}
}
\IEEEoverridecommandlockouts
\maketitle
\begin{abstract}
The design of systems implementing low precision neural networks with emerging memories such as resistive random access memory (RRAM) is a major lead for reducing the energy consumption of artificial intelligence (AI).
Multiple works have for example proposed in-memory architectures to implement low power binarized neural networks. These simple neural networks, where synaptic weights and neuronal activations assume binary values, can indeed approach state-of-the-art performance on vision tasks. In this work, we revisit one of these architectures where synapses are implemented in a differential fashion to reduce bit errors, and synaptic weights are read using precharge sense amplifiers. 
Based on experimental measurements on a hybrid  130~nm CMOS/RRAM chip and on circuit simulation, we show that the same memory array architecture can be used to implement ternary weights instead of binary weights, and that this technique is particularly appropriate if the sense amplifier is operated in near-threshold regime. We also show based on neural network simulation on the CIFAR-10 image recognition task that going from binary to ternary neural networks significantly increases neural network performance.
These results highlight that AI circuits function may sometimes be revisited when operated in low power regimes.
\end{abstract}


%
\IEEEpeerreviewmaketitle

\section{Introduction}

Artificial Intelligence has made tremendous progress in recent years due to the development of deep neural networks. 
Its deployment at the edge, however, is currently limited by the high power consumption of the associated algorithms \cite{xu2018scaling}.
Low precision neural networks are currently emerging as a solution, as they allow the development of low power consumption hardware specialized on deep learning inference \cite{hubara2017quantized}.
The most extreme case of low precision neural network, the Binarized Neural Network (BNN), also called XNOR-NET, is receiving special attention as it is  particularly efficient for hardware implementation: both synaptic weights and neuronal activations assume only binary values \cite{courbariaux2016binarized,rastegari2016xnor}.
Remarkably, this type of neural network can approach near-state-of-the-art performance on vision tasks \cite{lin2017towards}.
One particularly investigated lead is to fabricate hardware BNNs with emerging memories such as resistive RAM or memristors
\cite{bocquet2018,yu2016binary,giacomin2019robust,sun2018xnor,zhou2018new,natsui2018design,tang2017binary, lee2019adaptive}.
The low memory requirements of BNNs, as well as their reliance on simple arithmetic operations, make them indeed particularly adapted for ``in-memory'' or ``near-memory'' computing approaches, which achieve superior energy-efficiency by avoiding the von-Neumann bottleneck entirely.

In this work, we revisit one of this  hardware developed for the energy-efficient in-memory implementation of BNNs \cite{bocquet2018}, where the synaptic weights are implemented in a differential fashion.
We show that it can be extended to a more complex form of low precision neural networks, ternary neural network \cite{alemdar2017ternary} (TNN, also called Gated XNOR-NET, or GXNOR-NET \cite{deng2018gxnor}), where both synaptic weights and neuronal activations assume ternary values. 
The contribution of this work is as follows. After presenting the background of the work (section~\ref{sec:background}):

\begin{itemize}
\item We demonstrate experimentally on a fabricated 130~nm RRAM/CMOS hybrid chip a  strategy for implementing ternary weights using a precharge sense amplifier, which is particularly appropriate when the sense amplifier is operated in the near-threshold regime  (sec.~\ref{sec:circuit}).
\item We carry simulations to show the superiority of TNNs over BNNs on the canonical CIFAR-10 vision task, and evidence the error resilience of hardware TNNs (sec.~\ref{sec:network}).
\end{itemize}


\section{Background}
\label{sec:background}

The main equation in conventional neural networks is the computation of the neuronal activation $A_j =  f \left( \sum_i W_{ji}X_i \right),$ where $A_j$, the synaptic weights $W_{ji}$ and input neuronal activations $X_i$ assume real values, and $f$ is a non-linear activation function.
BNNs are a considerable simplification of conventional neural networks, in which all neuronal activations ($A_j$, $X_i$) and synaptic weights $W_{ji}$  can only take binary values meaning $+1$ and $-1$. 
Neuronal activation
then becomes:
\begin{equation}
\label{eq:activ_BNN}
    A_j = \mathrm{sign} \left( \sum_i  XNOR \left( W_{ji},X_i \right) -T_j \right),
\end{equation}
where $\mathrm{sign}$ is the sign function,   $T_j$ is a threshold associated with the neuron, and the $XNOR$ operation is defined in  Table~\ref{tab:gates}.
Training BNNs is a relatively sophisticated operation,  during which synapses need to be associated with a real value in addition to their binary value. Once training is finished, these real values can be discarded and the neural network is entirely binarized.
Due to their reduced memory requirements, and reliance on simple arithmetic operations, BNNs are especially appropriate for in- or near- memory implementations. 
In particular, multiple groups investigate the implementation of BNN inference with resistive memory tightly integrated at the core of CMOS \cite{bocquet2018,yu2016binary,giacomin2019robust,sun2018xnor,zhou2018new,natsui2018design,tang2017binary, lee2019adaptive}. Usually, resistive memory stores the synaptic weights $W_{ji}$.
However, this comes with a significant  challenge: resistive memory is prone to bit errors, and in digital applications, is typically used with strong error correcting codes (ECC). 
ECC, which requires large decoding circuit \cite{gregori2003chip}, goes against the principles of in- or near- memory computing.
For this reason,  \cite{bocquet2018} proposes a two-transistor/two-resistor (2T2R) structure, which reduces resistive memory bit errors, without the need of ECC decoding circuit, by storing synaptic weights in a differential fashion.
This allows for extremely efficient implementation of BNNs, and to use the resistive memory devices in very favorable programming conditions (low energy, high endurance).

In this work, we show that the same architecture may be used for a generalization of BNNs, TNNs
\footnote{
In the literature, the name  ``Ternary Neural Networks'' is sometimes  also used to refer to neural networks where the synaptic weights are ternarized, but the neuronal activations remains real or integer \cite{mellempudi2017ternary,nurvitadhi2017can}.}, 
where neuronal activations and synaptic weights
$A_j$, $X_i$ and $W_{ji}$ can now assume three  values: $+1$, $-1$ and $0$.
Equation~\eqref{eq:activ_BNN} now becomes:
\begin{equation}
\label{eq:activ_TNN}
    A_j = \phi \left( \sum_i  GXNOR \left( W_{ji},X_i \right) -T_j \right);
\end{equation}
$GXNOR$ is the ``gated'' XNOR operation that realizes the product between numbers with values $+1$, $-1$ and $0$ (Table~\ref{tab:gates}).
$\phi$ is an activation function that outputs $+1$ if its input is greater than a threshold $\Delta$, $-1$ if the input is lesser than $-\Delta$ and $0$ otherwise.
We show experimentally and by circuit simulation in sec.~\ref{sec:circuit} how the 2T2R BNN architecture can be extended to TNNs with practically no overhead, and in sec.~\ref{sec:network} the corresponding benefits in terms of neural network accuracy.

\begin{table}[tbp]
\caption{Truth Tables of the XNOR and GXNOR Gates}
\begin{center}
\begin{tabular}{|c|c|c|}
\hline
$W_{ji}$ & $X_i$ & $XNOR$  \\
\hline
$-1$ & $-1$ & $1$  \\
$-1$ & $1$ & $-1$  \\
$1$ & $-1$ & $-1$  \\
$1$ & $1$ & $1$  \\
\hline
\end{tabular}
\begin{tabular}{|c|c|c|}
\hline
$W_{ji}$ & $X_i$ & $GXNOR$  \\
\hline
$-1$ & $-1$ & $1$  \\
$-1$ & $1$ & $-1$  \\
$1$ & $-1$ & $-1$  \\
$1$ & $1$ & $1$  \\
$0$ & $X$ & $0$  \\
$X$ & $0$ & $0$  \\
\hline
\end{tabular}
\label{tab:gates}
\end{center}
\end{table}

\begin{figure}[ht]
	\centering
	\includegraphics[width=3.4in]{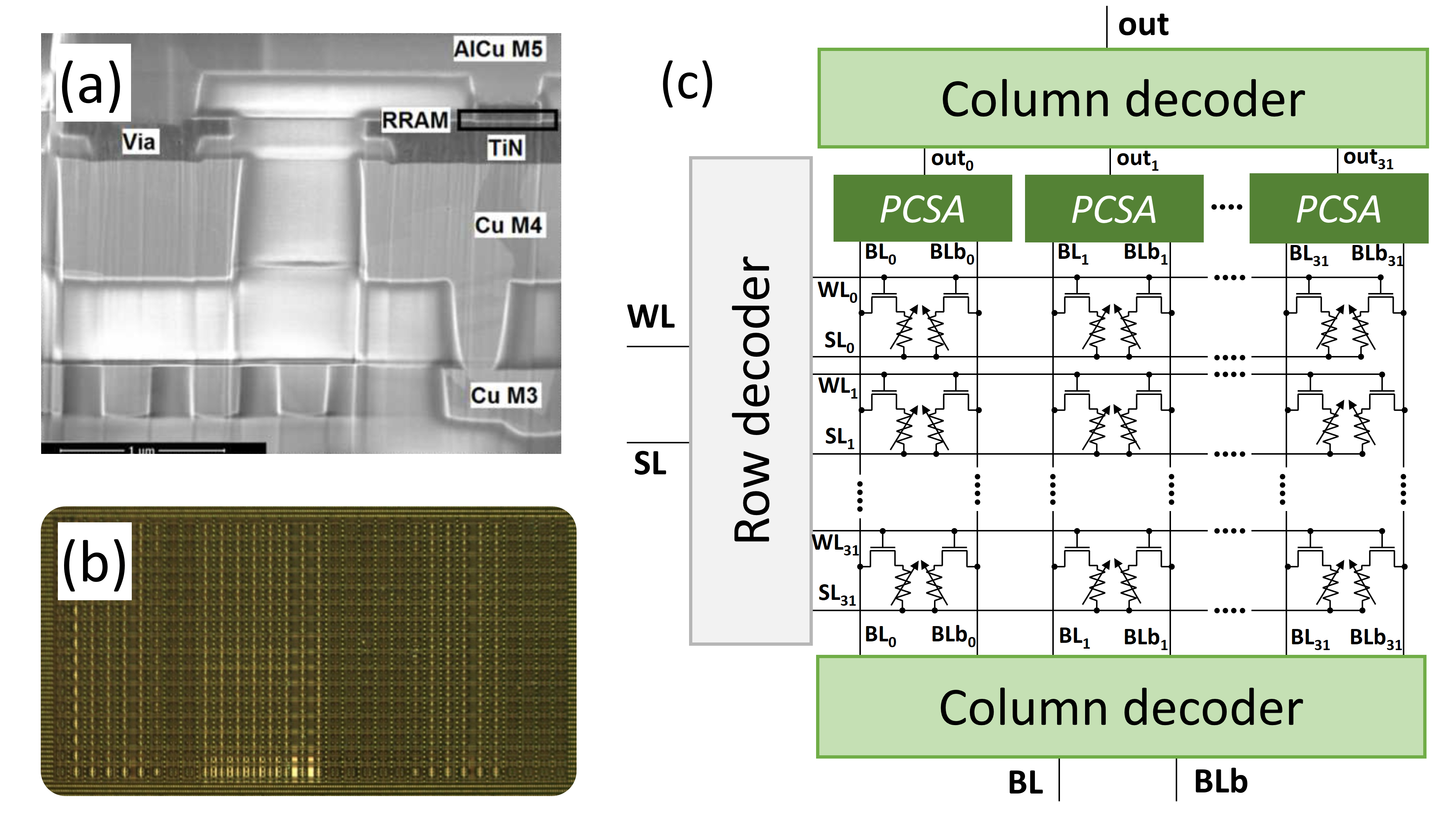}
	\caption{(a) Electron microscopy image of a hafnium oxide resistive memory cell (RRAM) integrated in the backend-of-line of a $130\nano\meter$ CMOS process. (b) Photograph and (c) simplified schematic of the hybrid CMOS/RRAM test chip characterized in this work. 	}
	\label{fig:testchip}
\end{figure}


\section{The Operation of A Precharge Sense Amplifier Can Provide Ternary Weights}
\label{sec:circuit}

\begin{figure}[htbp]
	\centering
	\includegraphics[width=2.5in]{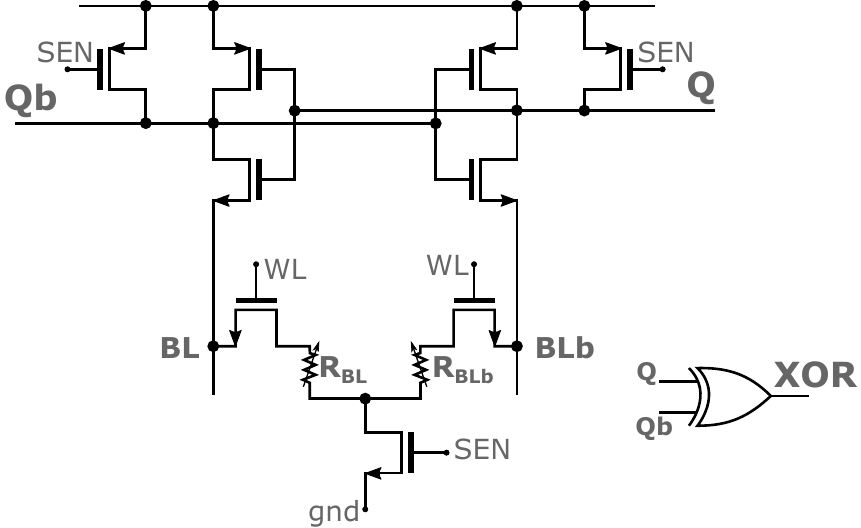}
	\caption{Schematic of the precharge sense amplifier fabricated in the test chip.}
	\label{fig:PCSA}
\end{figure}

\begin{figure}[htbp]
	\centering
	\includegraphics[width=\linewidth]{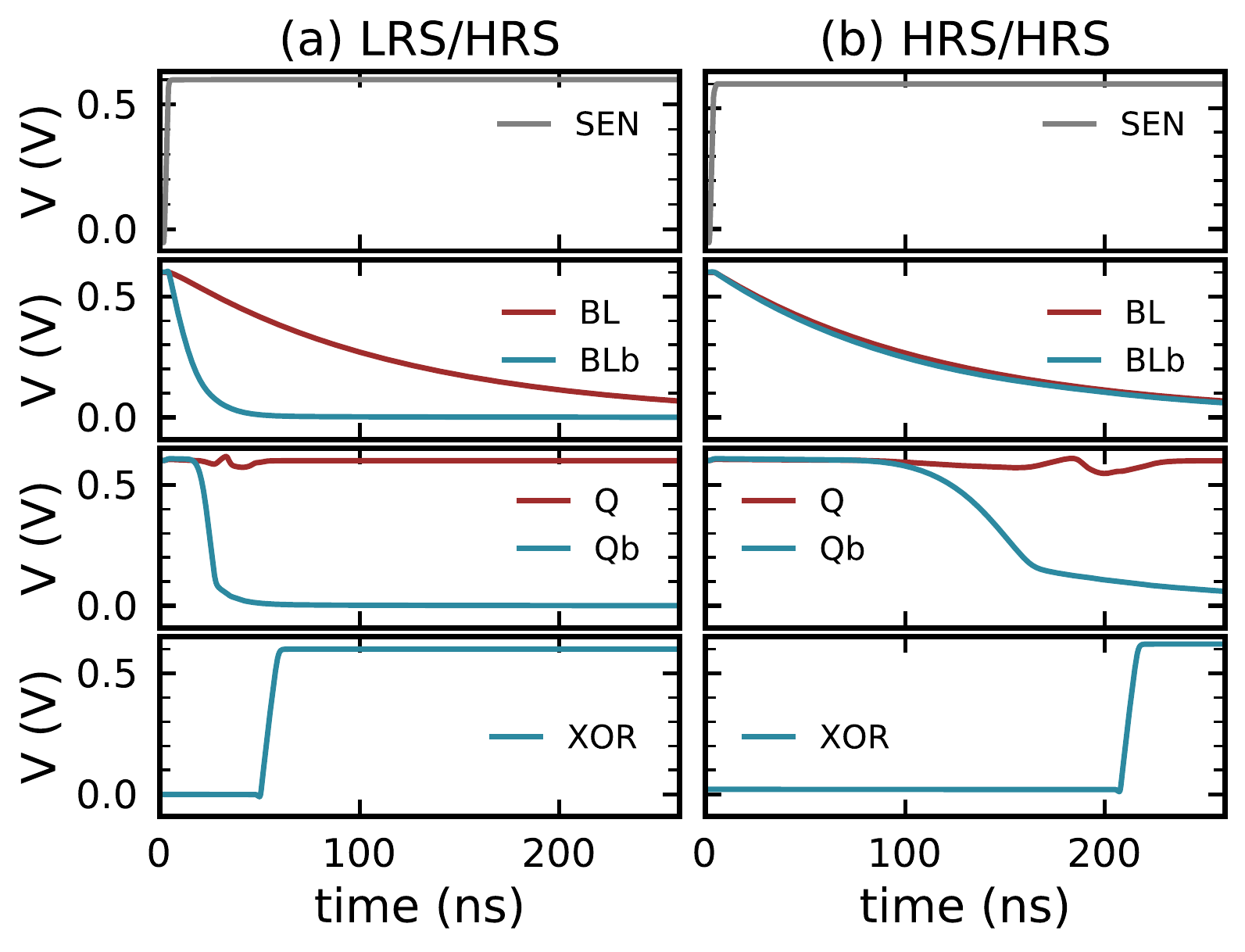}
	\caption{Circuit simulation of the precharge sense amplifier of Fig.~\ref{fig:PCSA} with a near-threshold supply voltage of $0.6\volt$, if the two devices are programmed in an (a) LRS / HRS ({$20\kilo\ohm$/$350\kilo\ohm$)} or (b) HRS/HRS ({$320\kilo\ohm$/$350\kilo\ohm$}) configuration.
	}
	\label{fig:SPICE}
\end{figure}
In this work, we use the architecture of \cite{bocquet2018}, where synaptic weights are stored in a differential fashion.
Each bit is implemented using two devices programmed either as low resistance state (LRS)/high resistance state (HRS) to mean weight $+1$ or HRS/LRS to mean weight $-1$. Fig.~\ref{fig:testchip} presents the test chip used for the experiments. 
This chip cointegrates $130\nano\meter$ CMOS and resistive memory in the back-end-of-line, between levels four and five of metal. The resistive memory cells are based on $10\nano\meter$ thick hafnium oxide (Fig.~\ref{fig:testchip}(a)).  Our experiments are based on a kilobit array incorporating all sense and periphery circuitry (Fig.~\ref{fig:testchip}(b-c)).
The synaptic weights are read using the onchip precharge sense amplifier (PCSA) presented in Fig.~\ref{fig:PCSA}, initially proposed  in \cite{zhao2009high}.
Fig.~\ref{fig:SPICE}(a) shows an electrical simulation of this circuit to explain its working principle.  These simulations are presented in the commercial $130\nano\meter$ technology used in our test chip, with a near-threshold supply voltage of $0.6 \volt$ \cite{nearth} (the nominal voltage in this process is $1.2\volt$).
In a first phase (SEN=0), the outputs Q and Qb are precharged to the supply voltage $V_{DD}$.
In the second phase (SEN=$V_{DD}$), each branch starts to discharge to the ground. The branch that has the resistive memory (BL or BLb) with lowest  electrical resistance discharges faster and causes its associated inverter to drive the output of the other inverter to the supply voltage. At the end of the process, the two outputs are therefore complementary, and can be used to tell which resistive memory has highest resistance and therefore the synaptic weight.
We observed that the convergence speed of a PCSA depends heavily on the resistance state of the two resistive memories. 
This effect is particularly magnified when the PCSA is used in near-threshold operation, 
 as presented here.
Fig.~\ref{fig:SPICE}(b) shows a simulation where the two devices BL and BLb were programmed in the HRS. We see that the two outputs converge to complementary values in $200\nano\second$, whereas only  $50 \nano\second$ were necessary in Fig.~\ref{fig:SPICE}(a), where the devices are programmed in complementary LRS/HRS states. 
These first simulations suggest a technique for implementing ternary weights using the 
memory array of our test chip. Similarly to when this array is used to implement BNN,  we propose to program the devices in the LRS/HRS configuration to mean the synaptic weight $1$, and HRS/LRS  to mean the synaptic weight $-1$. Additionally, we use the HRS/HRS configuration to mean synaptic $0$, while the LRS/LRS configuration is avoided.  The sense operation is performed during a duration of $70\nano\second$. If at the end of this period, outputs Q and Qb have differentiated and the output of the XOR gate is therefore 1, output Q determines the synaptic weight ($1$ or $-1$). Otherwise, output of the XOR gate is 0 and the weight is  determined to be $0$. 

\begin{figure}[tbp]	
\centering	
\includegraphics[width=2.7in]{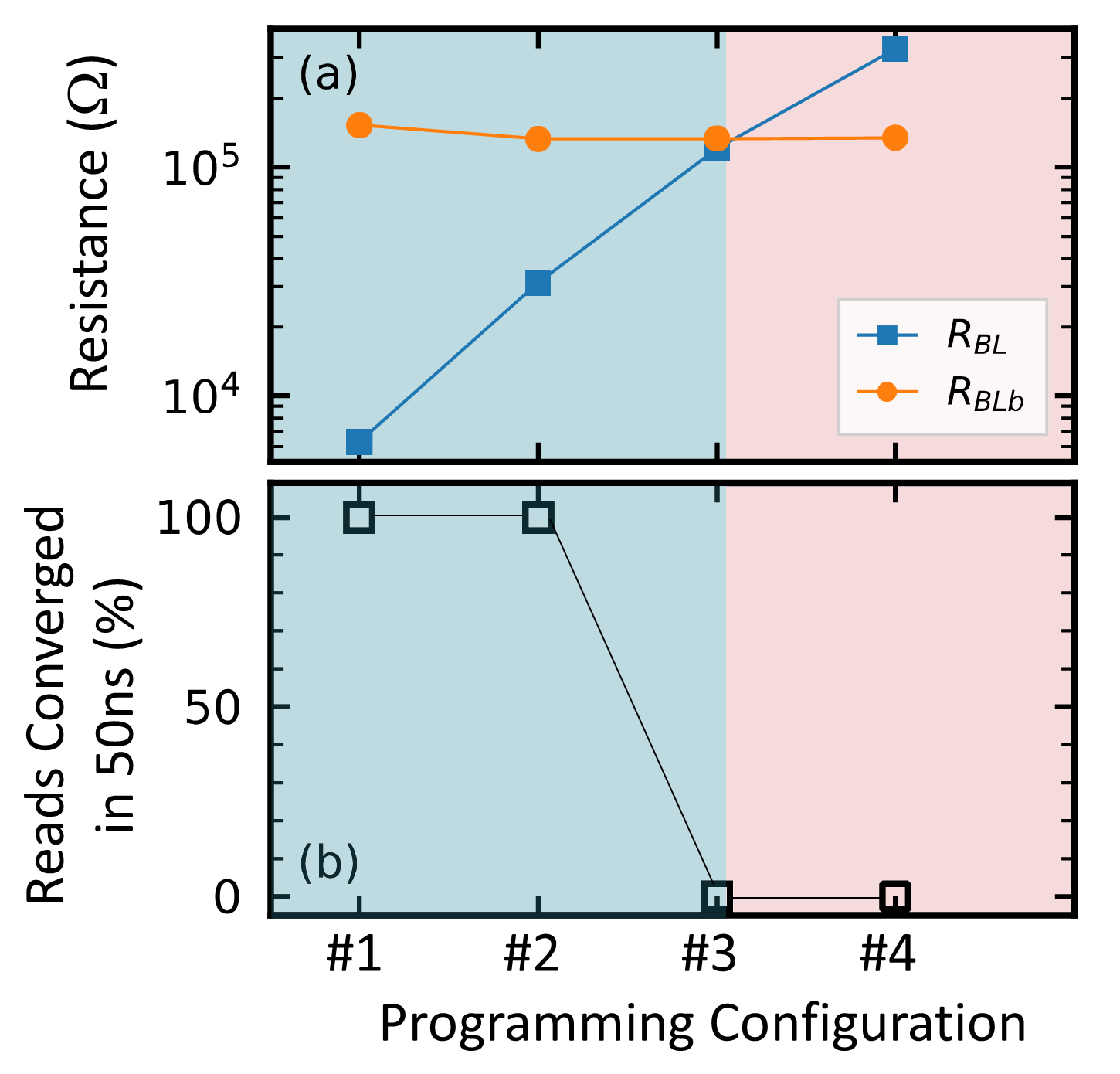} 
\caption{
Two devices have been programmed in four distinct programming conditions, presented in (a), and measured using an onchip sense amplifier. (b) Proportion of read operations that have converged in $50\nano\second$, over 100 trials.
} 	\label{fig:SenseVsR} \end{figure}

\begin{figure}[tbp]
	\centering 	
	\includegraphics[width=3.0in]{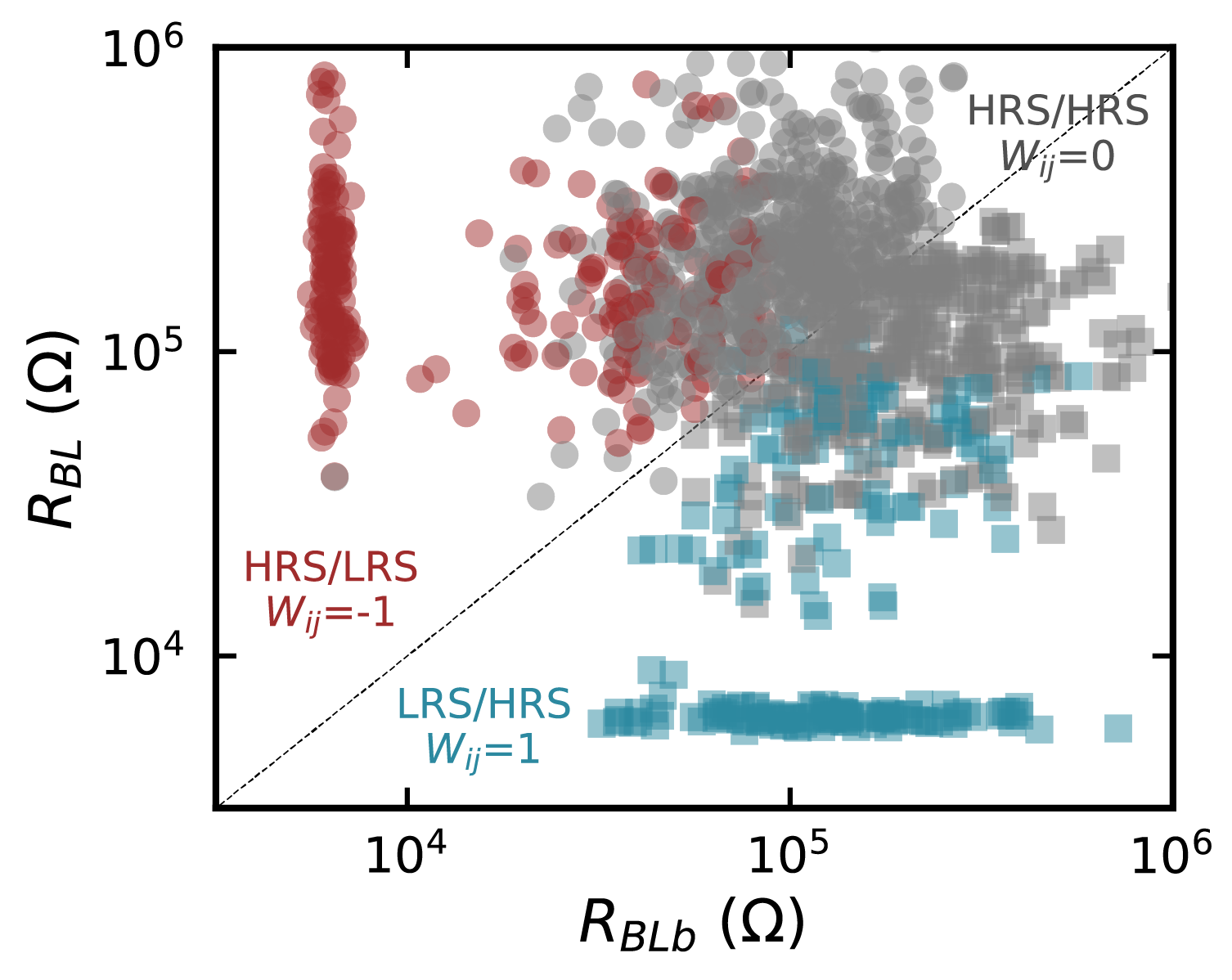} 
	\caption{For 109 device pairs programmed with multiple $R_{BL}/R_{BLb}$ configuration, value of the synaptic weight measured by the onchip sense amplifier using the  strategy described in  body text and $50\nano\second$ reading time.
	} 	\label{fig:multiniveau_2}
\end{figure}

\begin{figure}[bp]
	\centering
	\includegraphics[width=\linewidth]{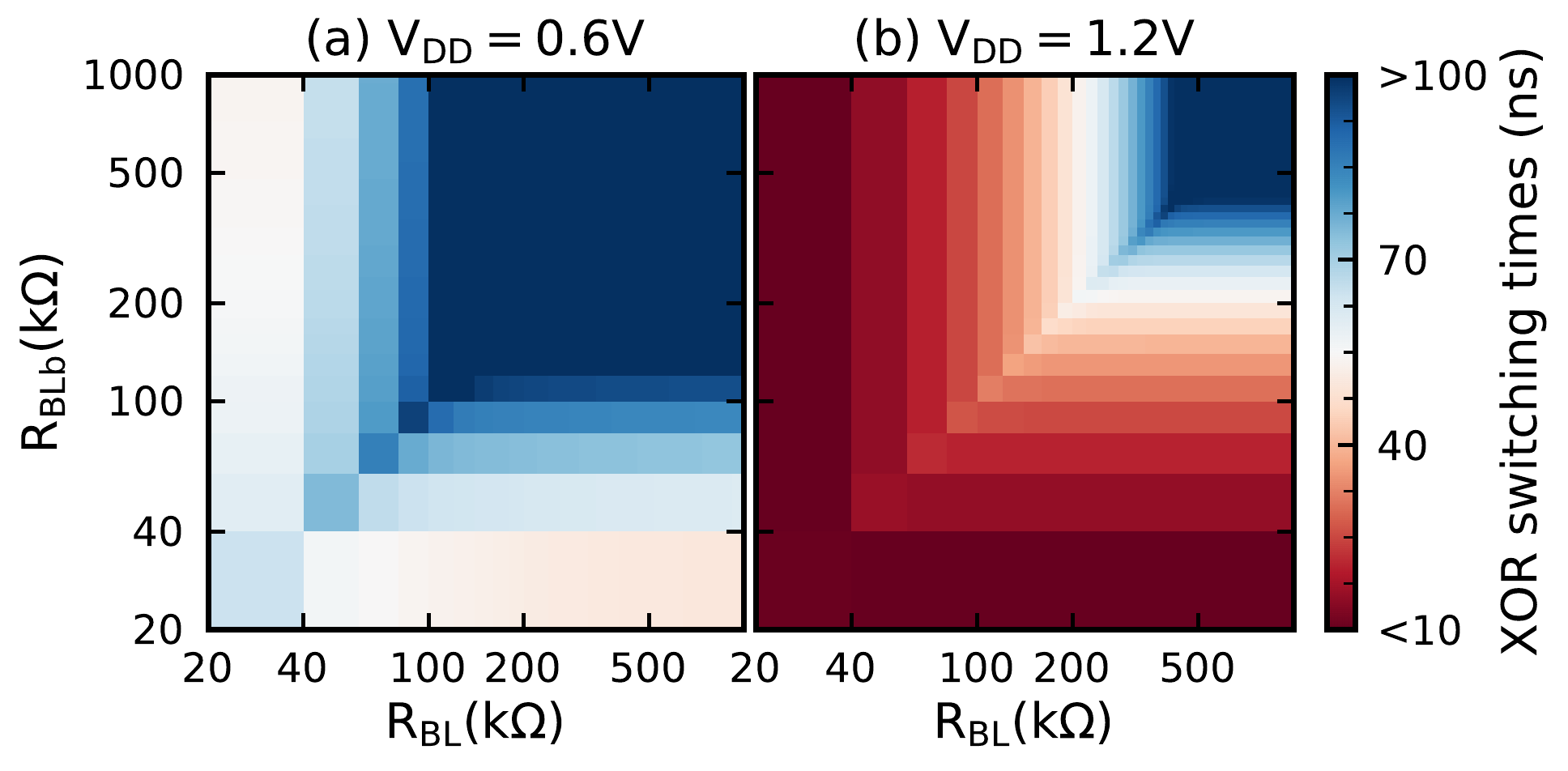}
	\caption{Switching time of a precharge sense amplifier, extracted from circuit simulations using the design kit of a $130\nano\meter$ commercial technology, as a function of the resistance of the BL and BLb complementary resistive memories. Simulations performed using (a) near-threshold $0.6\volt$ supply voltage and (b) nominal $1.2\volt$ supply voltage.}
	\label{fig:map}
\end{figure}

Experimental measurements on our 
test chip confirm that the PCSA can be used in this fashion. 
We first focus on one synapse of the memory array. We program one of the two devices (BLb) to a resistance of $10^5\ohm$. We then program its complementary device BL to several resistance values, and for each of them perform {100} read operations of duration $50\nano\second$, using an onchip PCSA operated in near-threshold.
Fig.~\ref{fig:SenseVsR}
plots the probability that the sense amplifier has converged during the read time.
In  $50\nano\second$, the read operation is only converged if the resistance of the BL device is significantly lower than $100\kilo\ohm$.
To evaluate this behavior in a wider range of programming conditions, we repeated the experiment on 109 devices and their complementary devices of the memory array programmed, each programmed 14 times, with various resistance values in the resistive memory, and performed a read operation  in $50\nano\second$ with an on-chip PCSA.
Fig.~\ref{fig:multiniveau_2} shows, for each couple of resistance value $R_{BL}$ and $R_{BLb}$ if the read operation was converged with $Q=V_{DD}$ (blue), meaning a weight of $1$, converged with $Q=0$ (red), meaning a weight of $-1$, or not converged (grey) meaning a weight of $0$. 
The results confirm that  LRS/HRS or HRS/LRS configurations may be used to mean weights $1$ and $-1$, and HRS/HRS for weight $0$. Relatively high values of the HRS should be targeted: the separation between the $1$ (or $-1$) and $0$ regions is not strict, and for intermediate resistance values we see that the read operation may or may not converge in $50ns$.

Extensive circuit simulations in the  $130\nano\meter$ technology of our test chip allow to evaluate this behavior in the general case.
Fig.~\ref{fig:map} shows the switching time of the PCSA as a function of the resistance of the two resistive memories BL and BLb, with nominal supply voltage ($1.2\volt$) and near-threshold supply voltage ($0.6\volt$). We see that for both supply voltages, the HRS/HRS configuration leads to longer switching times. 
In our technology, HRS states are typically characterized by resistances above $100\kilo\ohm$. We see that the operation in near-threshold exhibits a larger area of HRS/HRS values with a switching time above $70\nano\second$, corresponding to a 0 state. 
This implies a more robust detection of the 0 state in near threshold, compliant with the HRS variability.


\begin{figure}[bp]
	\centering
	\includegraphics[width=2.5in]{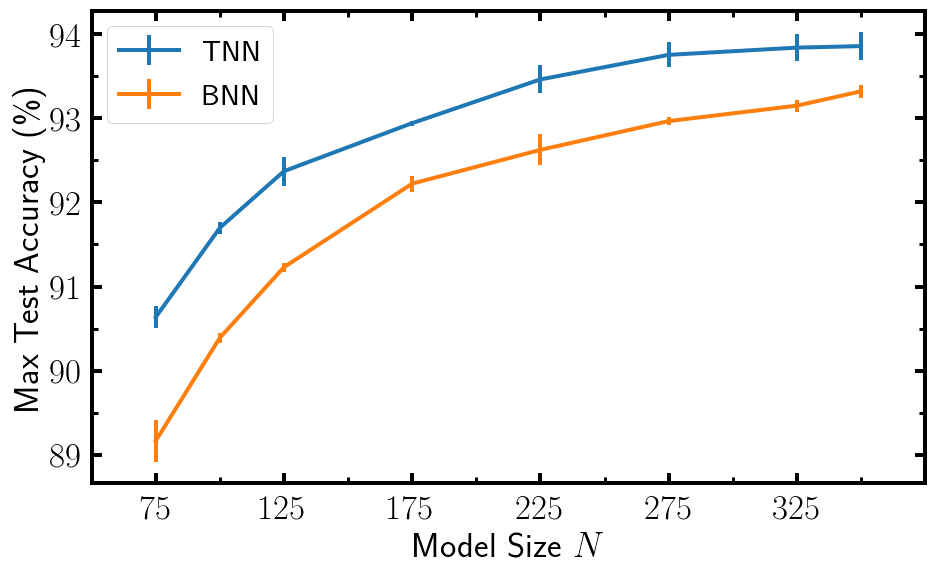}
	\caption{Maximum Test Accuracy reached during one training, averaged over five runs, for BNNs and TNNs with various model sizes on the CIFAR-10 dataset. Error bar is one standard deviation.}
	\label{fig:CIFAR10}
\end{figure}

\section{Network Level Benefits of Ternarization}
\label{sec:network}

We now investigate the accuracy gain when using ternarized instead of binarized networks.
We trained BNN and TNN versions of networks with VGG-type architectures  
\cite{simonyan2014very}
on the CIFAR-10 task of image recognition, consisting in classifying 
color images among ten classes. 
The architecture of our networks consists of six convolutional layers with kernel size three. The number of filters  at the first layer is called $N$ and is multiplied by two every two layers. Maximum-value pooling with kernel size two is used every two layers and batch-normalization \cite{ioffe2015batch} every layer. 
The classifier consists of one hidden layer of 512 units. For the TNN, the activation function has a threshold $\Delta = 5\cdot10^{-2}$ (as defined in section~\ref{sec:background}).
BNNs were trained following the methodology described in the Appendix of \cite{hirtzlin2019stochastic} and adapted from \cite{courbariaux2016binarized}. TNNs were trained with the methodology introduced in \cite{hubara2017quantized}. 
The training is performed using AdamW optimizer \cite{kingma2014adam, loshchilov2017fixing} with minibatch size $128$ and learning rate schedule used in \cite{loshchilov2016sgdr, loshchilov2017fixing}, resulting in a total of $700$ epochs. Data is augmented using random horizontal flip, and random choice between cropping after padding and random small rotations.
 

Fig.~\ref{fig:CIFAR10} shows the maximum test accuracies for different sizes of the model. 
TNNs always outperform BNNs with the same model size (and 
same number of synapses). The largest difference is seen for smaller model size, but a significant gap remains even for large models. 
Besides, the difference in the number of parameters required to reach a given accuracy for TNNs and BNNs increases with higher accuracies. There is therefore a clear advantage to use TNNs instead of BNNs. 

\begin{figure}[tbp]
	\centering
	\includegraphics[width=2.5in]{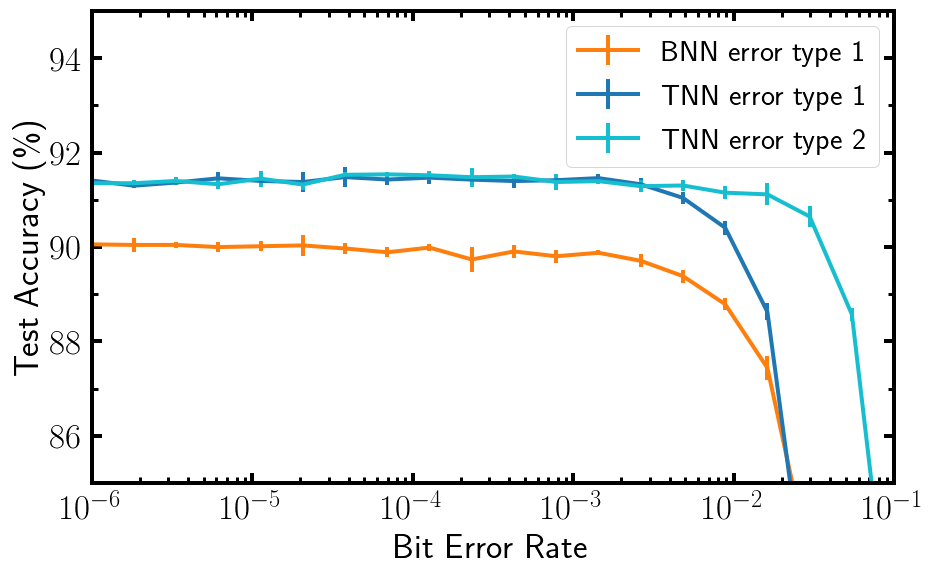}
	\caption{Impact of Bit Error Rate on the Test Accuracy at inference time for model size $N = 100$ TNN and BNN. Type 1 errors are sign switches (e.g. $+1$ mistaken for $-1$) and Type 2 errors are $0$ mistaken for $+1$ or $-1$, or $-1$ and $+1$ mistaken for $0$.}
	\label{fig:BT_errors}
\end{figure}

We then investigate the impact of bit errors in BNNs and TNNs to see if the advantage provided by TNNs remains when errors are taken into account. Two types of errors are investigated: Type 1 errors are sign switches,
while Type 2 errors are only defined for TNNs and correspond to $0$ mistaken for $+1$ or $-1$, and $+1$ or $-1$ mistaken for $0$. Type 1 errors are less likely than type 2 errors thanks to the differential reading scheme 2T2R, as seen in Fig.~\ref{fig:multiniveau_2}. 
Fig.~\ref{fig:BT_errors} shows the impact of  errors in the test accuracy for different values of the Bit error rate, averaged over five runs. 
Though Type 2 errors are more likely to occur with TNNs, their effect is not as serious as Type 1 errors as
the drop in accuracy for Type 2 errors occurs for bit error rates one order of magnitude higher than for Type 1 errors. Therefore TNNs still outperform BNNs when device imperfections are included.


\section{Conclusion}

In this work, we revisited a differential compute in-memory architecture for BNNs.
We showed experimentally on a hybrid CMOS/RRAM chip, and by circuit simulation that,
its sense amplifier is able to differentiate not only the LRS/HRS and HRS/LRS states, but also the HRS/HRS states. 
This allows the architecture to store ternary weights, and to provide a building block for TNNs. We showed by neural network simulation on the CIFAR-10 tasks the benefits of using ternary instead of binary networks, and the high resilience of TNNs to weights errors.
As this behavior is magnified in the slow but low power
near-threshold operation regime, our approach specially targets extremely energy-conscious applications such as uses within wireless sensors or medical applications. This work opens the way for increasing the edge intelligence in such contexts, and also highlights that near-threshold operations of circuits may sometimes provide opportunities for new functionalities.


\bibliography{IEEEabrv,Mabibliotheque}
\bibliographystyle{IEEEtran}

\end{document}